\documentstyle[aps,prb,multicol,epsfig,psfig,graphics]{revtex}
\tighten
\begin{document}

\title{Effective interactions in the stripe 
       state of the two-dimensional Hubbard model}

\author{S. Varlamov and G. Seibold}
\address{Institut f\"ur Physik, BTU Cottbus, PBox 101344, 
         03013 Cottbus, Germany}

\date{\today}
\maketitle

\begin{abstract}
We investigate the structure of the pairing potential in the stripe 
phase of the two-dimensional Hubbard model.
Based on the random phase approximation we discuss in detail the 
interactions in the charge- and spin channel and compare our calculations
with related considerations in commensurate antiferromagnets.
Our main finding is that due to the incommensurate charge-density
wave formation the exchange of collective modes in the charge channel
is significantly enhanced with respect to the spin bag approach whereas
due to the inhomogeneous charge distribution the coupling to transverse
spin fluctuations tends to be suppressed.
\end{abstract}


\vspace*{0.2cm}


\begin{multicols}{2}
\section{Introduction}

The existence of electronic inhomogeneities is now a well established 
experimental fact in underdoped high T$_c$ materials
(see e.g. \cite{HAMMEL,MESOT,TEITEL,BRINKMAN,MEHRING,HUNT}).
Especially in lanthanum
cuprates these inhomogeneities can develop in the form of antiphase
domain walls where quasi one-dimensional charge stripes are
separated by hole-free antiferromagnetically (AF) ordered regions. 
It is worth to note that stripe textures have been predicted as 
stable Hartree-Fock (HF) saddle-points of the Hubbard model \cite{Zaanen1,Rice}
before they were found experimentally in cuprates and nickelates.
Stripe textures in 
high-T$_c$ materials have been first detected in Nd-doped
lanthanum cuprate by neutron scattering experiments where due to the 
occurence of a low temperature tetragonal phase (LTT) both incommensurate 
AF and charge order are pinned \cite{TRAN}. The relevance of structural
distortions for the formation of antiphase domain wallls has been recently 
studied in Ref. \cite{NORMAND}.
In Nd-free compounds \cite{YAM,ARAI} these measurements  have up to now 
only detected the magnetic part of the scattering which remarkably displays
the same incommensurability than in the Nd-doped
system. In addition evidence for incommensurate charge fluctuations comes
from the analysis of optical phonon measurements in 
YBa$_2$Cu$_{3}$O$_{6.6}$ \cite{MOOK} and neutron diffraction experiments
\cite{BILLINGE} in La$_{2-x}$Sr$_x$CuO$_4$ (LSCO).
Taking the stripes as an experimental fact the question arises wether they 
have something to do with high-T$_c$ superconductivity or if they are just 
some 'strange byproduct' of the strong correlations which are 
at work in these materials.
In fact there are many experiments which support the point of view that 
stripe correlations could give rise to high transition temperatures. 
First , it has been shown that the stripe spacing in 
LSCO compounds 
is inversely proportional to the transition temperature up
to optimal doping \cite{YAM}. A close connection between transition 
temperature and stripe incommensurability has also been found in 
YBCO \cite{ARAI2} where interestingly there seems to appear a discontinuous
jump of incommensurability from $1/6$ at optimal doping to $1/8$ for
the T$_c$=60K underdoped compound.
Second, the occurence of domain walls can be interpreted in terms of a 
quantum critical point (QCP) \cite{CAST} located near optimal doping x$_{opt}$ 
which is a quite appealing concept in order to account for the 
anomalous normal state transport of high-T$_c$ materials around this 
particular doping. As a consequence the QCP scenario allows for a 
natural subdivision
of the phase diagram into overdoped metallic and the underdoped pseudogap
phase.
It has been argued \cite{TALLON} that the QCP can be deduced from 
the amplitude of the incommensurate magnetic peaks 
vansishing near x$_{opt}$. This coincides with experiments where 
superconducting order has been suppressed by strong pulsed 
magnetic fields \cite{BOEB} and which have revealed an underlying metal 
insulator transition at about the same concentration in agreement with the 
QCP concept. Temperature dependent measurements of the charge and spin 
sector in Nd-doped LSCO \cite{TRAN} as well as NQR experiments \cite{HUNT}
further support the 
idea that it is the charge rather than the spins which is reponsible
for the stripe state instability.

The possibility of d-wave pairing near a incommensurate CDW instability
has been investigated in Ref. \cite{PERALI} within a BCS-type scheme.
This analysis was restricted to the overdoped and optimally doped
region of the phase diagram
where according to the QCP scenario strong ICDW fluctuations are present but
without symmetry breaking of the translational invariance.
In the underdoped system the opening of a ICDW gap naturally leads
to a suppression of the superconducting order parameter due to
a reduction in the density of states around the Fermi level (see e.g.
\cite{BALSEIRO},\cite{LAR}). However, in addition the ICDW fluctuations
are strongly modified in the symmetry-broken state but
to our knowledge no detailed analysis of the corresponding pairing 
interaction has been performed yet. 
The present paper is dedicated to explore the effective electron-electron
interactions in the underdoped regime of the phase diagram thus starting
with a ground state which exhibits long-range incommensurate stripe
order. Our considerations are based on the Hartree-Fock (HF) decoupled
two-dimensional Hubbard model where we obtain the domain wall structure 
via a self-consistent iteration of the on-site charge and spin expectation
values.  
From the formal point of view our investigations can be viewed as a 
generalization of the spin-bag approach \cite{SCHRIEFF} to incommensurate
antiferromagnetism. The basic idea of the spin-bag mechansim is that 
a hole doped into the commensurate antiferromagnet locally perturbes
the AF order parameter and consequently two holes gain energy when
they share a common deformation (bag) in analogy to a standard lattice
bipolaron. In Ref. \cite{SCHRIEFF} this scenario has been worked out
by considering the pairing in the longitudinal spin channel of a
commensurate antiferromagnet which indeed turns out to be attractive
for small momentum transfers whereas it becomes repulsive for
large ${\bf q} \sim (\pi,\pi)$. The charge channel plays a minor role 
for pairing in the spin-bag approach since it does not exhibit an 
instability and renormalization effects are rather weak.
It has been argued in Ref. \cite{SCHRIEFF}
that the repulsion in the transverse spin channel at large ${\bf q}$ may
be suppressed by the coherence factors entering the vertex of the
pairing interaction. However, as pointed out by Frenkel and Hanke 
\cite{FRENKEL} the vanishing of the vertex is cancelled by the 
Goldstone pole in the spin susceptibility thus leading to a finite
repulsive contribution in the transverse channel for momenta 
${\bf q} \sim (\pi,\pi)$ comparable to that for the amplitude fluctuations.

We want to emphasize that our HF approach to the stripe phase
fails to account for some of the
experimental facts mentioned above but on the other hand allows for a 
simple description of some basic features regarding the electron-electron
interaction in a system posessing stripe order.
First, it is well known \cite{ZAANEN} that a HF decoupling of the one-band
Hubbard model cannot properly describe the observed stripe charge structure
in the cuprates which is characterized by one hole for every second 
unit cell along the domain wall ('half-filled'). 
In contrast HF calculations result in completely filled stripes thus
predicting twice the observed stripe periodicity and it has turned
out that correlations beyond HF theory have to be included 
\cite{goetz1,fleck} in order to account for the measured stripe structure
within the Hubbard model.

Thus in our investigations we will restrict to excitations perpendicular
to the domain walls  which should be less affected by the actual stripe
filling than excitations along the walls which of course crucially depend
on the doping of the stripes.
Second, it has been shown in Ref. \cite{SCHULTZ} that the 
HF stripe instability is due to an instability in the spin channel and the 
corresponding charge modulation
is generated as a higher harmonic of the longitudinal spin scattering.
Thus in our model the spins are the driving force behind the domain
wall formation in contrast to what is observed in Nd-doped LSCO.
An alternative approach where antiphase domain walls are enslaved by
a frustrated phase seperation scenario has been discussed in \cite{ZACHAR}.
However, since we restrict on zero temperature we believe that within the
limitations discussed above our approach  can capture some features of the 
electron-electron  interaction in the stripe state of underdoped cuprates. 
We note that a similar study for the calculation of correlation functions 
has been performed in Ref. \cite{KANESHITA} where the authors focussed 
on the renormalization of the spin velocity in the stripe state. 

The paper is organized as follows. In Sec. II we outline the formalism
of our approach starting from HF theory and the subsequent
derivation of effective hamiltonians for the residual RPA-type interactions
in the stripe phase.
In Sec. III we present results for these interactions focussing
on the different contributions from vertex functions and susceptibilities
respectively.
We finally conclude our discussion in Sec. IV.

\section{Formalism}

\subsection{Mean-field theory}

We start with the Hubbard hamiltonian

\begin{equation}
H=\sum_{k,\sigma}\epsilon_{k} c_{k,\sigma}^{\dagger}c_{k,\sigma}
 + \frac{U}{2}\sum_{q,\sigma} \rho_{-q,-\sigma} \rho_{q,\sigma} 
\end{equation}

where $c_{k,\sigma}^{(\dagger)}$ destroys (creates) an electron in the
state k and energy $\epsilon_k$, 
$\rho_{q,\sigma}=\sum_{k} c_{k+q,\sigma}^{\dagger}c_{k,\sigma}$ 
denotes the density operator. 

For one elecron per site the ground state shows two-sublattice commensurate
antiferromagnetic order. It has been shown by
Schulz \cite{SCHULTZ} using the RPA-type Stoner criterion that for a small 
number of holes the system
posesses a magnetic instability at an incommensurate modulation
wavevector $(\pi,\pi)+{\bf q_s}$. In principle the corresponding (mean-field)
symmetry-broken state could be realized either by a modulation of the
transverse spin degrees of freedom i.e. a spiral phase \cite{ARRIGONI}
or the corresponding modulation of the longitudinal magnetization.
From general arguments \cite{OVERHAUSER} and also from unrestricted 
Hartree-Fock calculations (see e.g. \cite{ZAANEN}) one finds that usually
the symmetry breaking occurs in the longitudinal channel of the
magnetic system. As a consequence this leads to a strong coupling between 
charge- and spin degrees of freedom resulting in the formation of
antiphase domain walls where the charge is distributed along the phase
boundaries. It should be noted that the inclusion of the vacuum 
renormalization of the effective interaction may stabilize the commensurate
antiferromagnet at small dopings \cite{CHUBUKOV}. 

Within unrestricted HF theory (see e.g. \cite{INUI}) one finds
that for small values of the on-site interaction U the stripes 
are oriented along the verticals and whereas for large U they
run along the (1,1)-direction. In the following we consider a rather
small value of U close to the instability in order to realize the
vertical stripe phase. 

Moreover we restrict ourselves to an energy dispersion arising from nearest 
neighbor hopping $\varepsilon_k=-2t(cos\,k_x+cos\,k_y)$ with bandwidth
$B=8t=4eV$. 
For doping $\delta=0.13$ (measured from half-filling) 
one finds from the Stoner criterion the occurence of the magnetic instability 
at ${\bf Q^t}=\pi(1 \pm 1/8, 1)$ and ${\bf Q^t}=\pi(1,1\pm 1/8)$
when the Hubbard interaction exceeds the critical value U$_c \sim 0.9eV$.
In principle these four equivalent wave vectors allow either for an
unidirectional stripe structure or for a two-dimensional grid pattern 
of antiphase domain walls. In the spin glass phase of LSCO where the
stripes run along the diagonals neutron scattering provides convincing
evidence for a one-dimensional modulation \cite{MATSUDA}, however, 
in the superconducting regime this point remains unclear although from
general arguments \cite{TRANQUADA} unidirectional scattering 
seems to be favored.

Restricting ourselves to the (longitudinal) domain wall structure 
the Hubbard interaction can be decoupled via

\begin{eqnarray}\label{2}
\rho_{-q,-\sigma} \rho_{q,\sigma}  &\approx&  \frac{1}{4} \left(
\langle \rho_{-q} \rangle \rho_q + \langle \rho_{q} \rangle \rho_{-q} 
- \langle \mu_{-q} \rangle \mu_{q} 
- \langle \mu_{q} \rangle \mu_{-q}\right)            \nonumber \\
&-& \frac{1}{4} \left( \langle \rho_{q} \rangle \langle \rho_{-q} \rangle
- \langle \mu_{q} \rangle \langle \mu_{-q} \rangle \right)
\end{eqnarray}

where
$\langle \rho_q \rangle=\sum_\sigma \langle \rho_{q,\sigma}\rangle$ and
$\langle \mu_q \rangle=\sum_\sigma \sigma \,\langle \rho_{q,\sigma}\rangle$
denote the charge and spin densities respectively.
Although the Stoner criterion signals the instability at a single
wave vector ${\bf Q^t}=(\pi,\pi)+{\bf q_s}$ the anharmonicity of the 
domain wall structure in the symmetry broken phase naturally 
incorporates higher harmonics also. It is well known \cite{Zachar0}
that the topological nature of the stripe phase implies the relation
${\bf q_c}=2{\bf q_s}$ between the basic charge (${\bf q_c}$) and spin
(${\bf q_s}$) modulations.

In order to illustrate the role of these harmonics in the charge-
and spin channel respectively, we have sketched in Fig. 1 the 
charge ($\chi(x)$) and spin order ($\Delta(x)$) parameter for a simple 
rectangular stripe pattern in the periodicity interval $[0,l]$.
One can easily convince oneself that the corresponding Fourier coefficients 
$\chi_k$ and $\Delta_k$ are given by

\begin{eqnarray}
\chi_k=\frac{4\chi_0}{k \pi}\cos(k\pi/2)\cos(\frac{k\pi \xi}{l}) \nonumber \\
\Delta_k=\frac{4\Delta_0}{k \pi}\sin(k\pi/2)\sin(\frac{k\pi \xi}{l})
\end{eqnarray}

where $\xi$ denotes the width of the magnetic domains.
Thus the magnetic scattering only involves odd harmonics whereas the 
charge scattering is due to the even harmonics. 
This fact holds in general when both magnetic and charge stripes are symmetric
under reflection with respect to their central axis. 
As a result one can describe the modulation 
of charge and spin densities in terms of the single wave vector ${\bf Q^t}$ 

\begin{eqnarray}\label{3}
\langle\rho({\bf r})\rangle&=&2\sum_{p(even)=2}^{M-1} \langle\rho_{p{\bf Q^t}}
\rangle \cos(p{\bf Q^t r}),\nonumber\\
\langle \mu({\bf r})\rangle&=& 2 \sum_{p(odd)=1}^{M-1} 
\langle \mu_{p{\bf Q^t}}\rangle \sin(p{\bf Q^t r})
\end{eqnarray}

where the commensurability M of the stripe is defined as the number of 
scattering
events in the same direction needed to return to the equivalent state, i.e.
$|{\bf k}+M{\bf Q^{t}}\rangle=|{\bf k}\rangle$. Note that 
within this definition a large
commensurability M not necessarily implies a scattering vector 
${\bf Q^t}$ very distant from $(\pi,\pi)$.

\begin{figure}
\begin{center}
\hspace{6cm}{{\psfig{figure=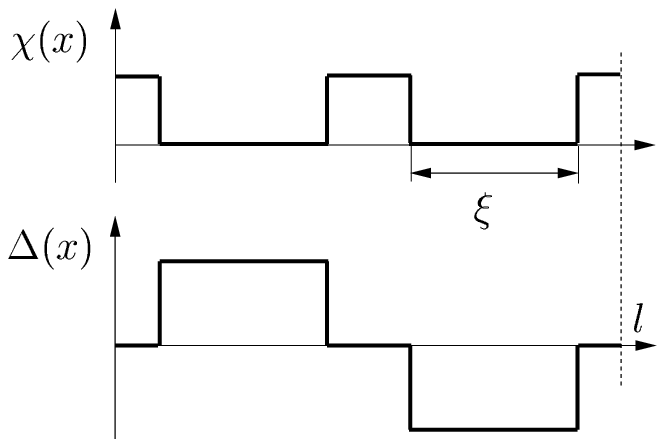,width=7cm}}}
\end{center}
\vspace*{0.2cm}
{\small FIG.1. Illustrative plot for the charge ($\chi(x)$) and 
spin order ($\Delta(x)$) parameter of a rectangular stripe pattern. }
\end{figure}

For the  vertical stripe orientations along
the x-direction considered here we can write 
${\bf Q^t}=\pi(\frac{2}{M}+\Theta(M-2),1)$ where the step function
is defined with $\Theta(0)=0$. This guarantees the correct behavior 
for commensurate AF (M=2) i.e. ${\bf Q^t}=(\pi,\pi)$. Note that within
these definitions the commensurability M corresponds to the period of
the stripe structure (in units of the lattice constant) as is indicated in
Fig. 2a. Furthermore we are now able to construct the reduced Brillouin 
zone (BZ) which is spanned by 
vectors $\bf{Q}^t$ and ${\bf Q}_{\perp}=\frac{4\pi}{(2+M\Theta(M-2))^2+M^2}
(-M,2+M\Theta(M-2))$. 
In order to illustrate the new zone scheme we plot in Fig. 2b the reduced 
BZ for the case $M=16$ which will be studied below. 
Note that the reduced BZ is {\bf not} only
due to charge order (the corresponding vertically striped BZ is also 
indicated in Fig. 2b) but is rotated and divided in half due to 
the presence of (modulated) antiferromagnetism. 

Introducing reduced zone notation for the electron operators 

\begin{equation}
c_n(k,\sigma)\equiv c_{k_x+(n-1)Q_x^{t}, k_y+(n-1)Q_y^{t}, \sigma} 
\end{equation}

we thus obtain the following hamiltonian

\begin{eqnarray}
H &=& \sum_{n,k,\sigma}\epsilon_{n}(k) c_n^{\dagger}(k,\sigma)c_n(k,\sigma)  
\nonumber \\
&+& \sum_{p,n,k,\sigma} \lbrack 
\chi_{2p}  c_{n+2p}^{\dagger}  (k,\sigma)c_n(k,\sigma)
+ h.c. \rbrack  \nonumber\\
&+& \sum_{p,n,k,\sigma} \sigma \lbrack 
  \Delta_p   c_{n+p}^{\dagger} (k,\sigma)c_n(k,\sigma)
+ h.c. \rbrack \nonumber \\
&+&\frac{2N}{U}\sum_p (|\Delta_{p}|^2-|\chi_{2p}|^2) \label{h1}
\end{eqnarray}

where we have introduced the charge and spin order parameters 
as $\chi_{2p}=(U/2)\langle \rho_{2p{\bf Q^t}}\rangle$ and 
$\Delta_p=-(U/2)\langle\mu_{p{\bf Q^t}}\rangle$.

The hamiltonian eq. (\ref{h1}) can be diagonalized via

\begin{equation}\label{Trans}
c_n(k,\sigma)=\sum_{m=1}^M A_{nm}(k,\sigma) f_m(k,\sigma) 
\end{equation}

\begin{figure}
\begin{center}
\hspace{6cm}{{\psfig{figure=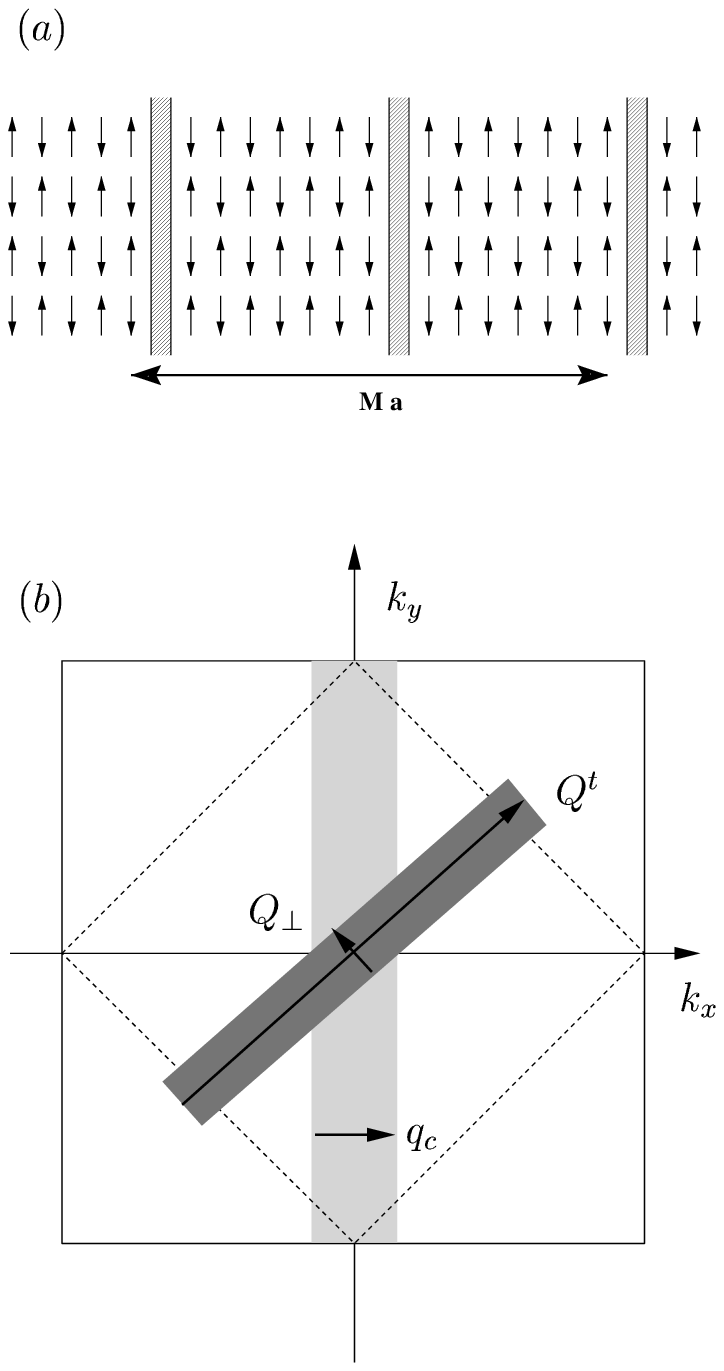,width=7cm}}}
\end{center}
\vspace*{0.2cm}
{\small FIG.2 a) Sketch of a vertical stripe pattern where the shaded areas 
correspond to the charged domain walls. The commensurability M times
the lattice constant a defines the periodicity of the structure.
b) The large square represents the full Brillouin zone whereas
the dark shaded rectangle depicts the reduced Brillouin zone for 
incommensurability $M=16$. The latter can be constructed from the
reduced zone for charge order only (light shaded rectangle) which is 
rotated and divided in half due to antiferromagnetism. Since also
the AF order is modulated the boundaries of the dark shaded rectangle 
are not parallel to those of the reduced zone for commensurate
antiferromagnetism (dashed line). }
\end{figure}

and from the symmetry of H it turns out that the transformation matrices 
obey the relations
$A_{nm}(k) \equiv A_{nm}(k,\sigma)=(-1)^n A_{nm}(k,-\sigma)$ and 
$A_{nm}(k,\sigma)=A^{\ast}_{M-n+2,M-m+2}(-k,\sigma)$.

We finally obtain the hamiltonian in diagonal representation as

\begin{eqnarray}\label{mfham}
H &=&\sum_{k,\sigma}\sum_{n}E_n(k)f_n^{\dagger}
(k,\sigma)f_n(k,\sigma) \nonumber \\
&+&\frac{2N}{U}\sum_p (|\Delta_{p}|^2-|\chi_{2p}|^2)
\end{eqnarray}

where $E_n(k)$ denotes the dispersion of the n-th subband in the reduced BZ
(dark shaded rectangle in Fig. 2b). The order parameters have to be determined
self-consistently.

In Fig. 3 we show a cut of the bandstructure through the reduced zone
and the density
of states for a stripe structure with commensurability M=16 corresponding
to the scattering vector ${\bf Q^t}$ as defined above.
As can be seen the spectrum is composed of 16 quasi one-dimensional subbands. 
Since the choosen carrier concentration $\delta=0.13$ slightly exceeds 
the integer fraction $\delta=1/8$ the chemical potential is located
within but near the top edge of the 7th subband. 
 
The following sections deal with the RPA fluctuations around the HF
stripe state and it is quite instructive to consider the functional
dependence of the ground state energy from the spin- and charge
order parameters which is shown in Fig. 4. 
In fact, since RPA excitations can be derived from the 
expansion around the HF saddle-point in the density fluctuations 
(see e.g. \cite{BLAIZOT})
one can already qualitatively deduce from Fig. 4 the possible instabilities 
of the system.
Note that for the $E(\Delta_1)$ curve the charge order parameter
is adjusted selfconsistently while for the $E(\chi_2$) curve the spin order
parameter is taken at the saddle point. 
The $(\langle H \rangle, \Delta_1)$ curve displays the standard behavior
of a second-order phase transition towards the SDW ground state with minimum
at  $\Delta_1=0.14 eV$. On the other hand it turns out that the saddle-point
of  $(\langle H \rangle, \chi_2)$
corresponds to a maximum in the energy-order parameter space
which is also obvious from the 
decoupling eq. (\ref{2}) \cite{foot1}. 
This simple picture already
demonstrates that in the present framework charge order is completely
due to the SDW instability. 

\begin{figure}
\begin{center}
\hspace{6cm}{{\psfig{figure=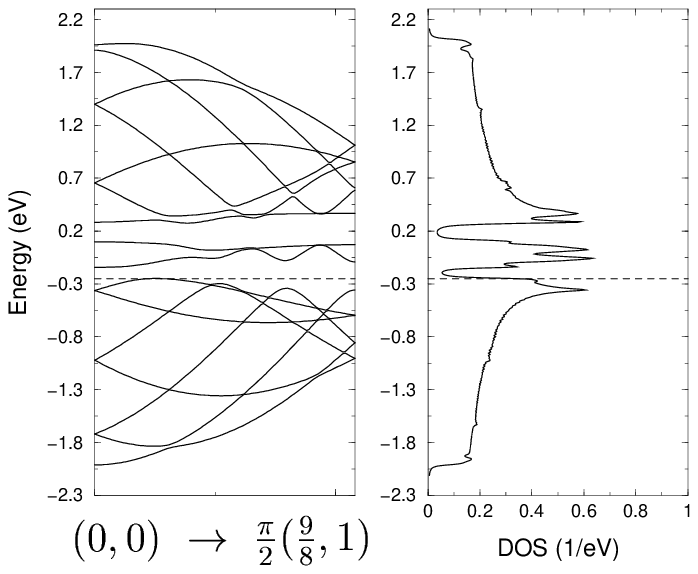,width=7.5cm}}}
\end{center}
\vspace*{0.2cm}
{\small FIG.3 Band structure and density of states for a stripe array with 
commensurability M=16 and U/t=1.3. The wave vector cut of the bandstructure
is from $(0,0)$ along ${\bf Q^{t}}$ (see Fig. 2b).  
The dashed line indicates the position of the chemical potential. }
\end{figure}

\begin{figure}
\begin{center}
\hspace{6cm}{{\psfig{figure=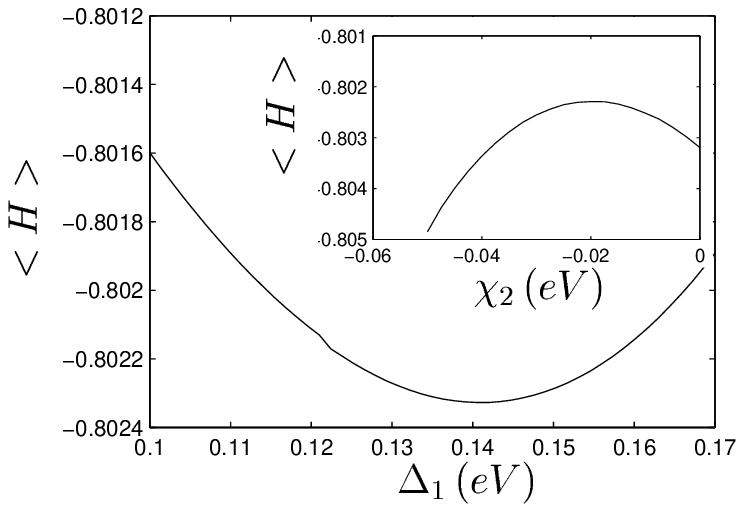,width=7cm}}}
\end{center}
\vspace*{0.2cm}
{\small FIG.4 Free energy as a function of the SDW order parameter $\Delta_1$.
The inset displays the free energy as a function of the CDW order parameter $\chi_2$.}
\end{figure}

Moreover, note that for $\chi_2 \rightarrow 0$
the $(\langle H \rangle, \chi_2)$ curve has a finite derivative which signals
the absence of critical fluctuations in the charge channel for the
homogeneous system in contrast to models where the instability mechanism
is due to frustrated phase separation \cite{EMERY,CAST}.

\subsection{Susceptibilities for general commensurability}

We start from the following definitions for charge and spin-density 
correlations functions

\begin{eqnarray}
\chi^{+-}_{mm'}(q,q',t)&=&\frac{i}{N}
\langle TS_{q+(m-1)Q^t}^+(t) S_{-q'-(m'-1)Q^t}^-(0)\rangle,  \nonumber \\
\chi^{\sigma\sigma'}_{mm'}(q,q',t)&=&\frac{i}{N}
\langle T \rho_{q+(m-1)Q^t,\sigma}(t) \rho_{-q'-(m'-1)Q^t,\sigma'}(0)\rangle,  
\nonumber \\
\chi^{00}_{mm'}(q,q',t)&=&\sum_{\sigma\sigma'}\chi^{\sigma\sigma'}_{mm'}(q,q',t),
\nonumber \\
\chi^{zz}_{mm'}(q,q',t)&=&\sum_{\sigma\sigma'}\sigma\sigma'
\chi^{\sigma\sigma'}_{mm'}(q,q',t), \label{SUS}
\end{eqnarray}

where the wave vectors $q$ are within the reduced zone. 
Since the mean-field hamiltonian eq. (\ref{mfham}) commutes with
$\sum_i S_i^z$ the excitations in the longitudinal spin and charge channel
are decoupled from the transverse channel where the solution of the 
RPA approximated Dyson equation for the latter case leads to 

\begin{equation}
\chi_{m\,m'}^{+-}(q,\omega)=[1-U\chi_{m\,m'}^{(0),+-}(q,\omega)]^{-1}
\chi_{m\,m''}^{(0),+-}(q,\omega)
\end{equation}
 
The bare transverse stripe susceptibilities read as

\begin{eqnarray}
&&\chi_{m\,m'}^{(0),+-}(q,\omega)=\frac{1}{N}\sum_{k\,s\,t}
\Gamma_{s\,t\,m}^+(\widetilde{k+q},k)\,
\Gamma_{t\,s\,m'}^-(k,\widetilde{k+q}) \nonumber\\
&&[\frac{n_t(\widetilde{k})[1-n_s(\widetilde{k+q})]}
{\omega+E_s(\widetilde{k+q})-E_t(k)-i\eta}
-\frac{n_s(\widetilde{k+q})[1-n_t(k)]}
{\omega+E_s(\widetilde{{k+q}})-E_t(k)+i\eta}] \label{chipm}
\end{eqnarray}

where we have introduced the vertex functions:

\begin{eqnarray}
&&\Gamma_{stm}^{+}(\widetilde{k+q},k)= \sum_n A^{*}_{{\cal N},s}
(\widetilde{k+q})A_{n,t}(k)(-1)^{n-1}  \nonumber\\
&&\Gamma_{tsm}^{-}(k,\widetilde{k+q})=(\Gamma_{stm}^{+}(\widetilde{k+q},k))^*
\end{eqnarray}

and $A_{nt}(k)$ is defined in Eq. (\ref{Trans}).
Here the index ${\cal N}={\cal N}(n,m,k,q)$ defines the number of the
BZ which contains the unreduced state $|k+q\rangle$ and $\widetilde{k+q}$
denotes the corresponding reduced wave vector.

We now turn to the evaluation of the correlation function in the
longitudinal channel. In this case we start from the RPA decoupled Dyson 
equation for $\chi_{m\,m'}^{\sigma\,\sigma'}(q,\omega)$

\begin{equation}\label{sz}
\chi_{m\,m'}^{\sigma\,\sigma'}(q,\omega)=\chi_{m\,m'}^{(0),\sigma\,\sigma}
(q,\omega)
+U\chi_{m\,m''}^{(0),\sigma\,\sigma}(q,\omega)
\chi_{m''\,m}^{-\sigma\,\sigma'}(q,\omega)
\end{equation}

where

\begin{eqnarray}
&&\chi_{m\,m'}^{(0),\sigma\,\sigma}(q,\omega)=\frac{1}{N}\sum_{k\,s\,t}
\Gamma_{s\,t\,m}^\sigma(\widetilde{k+q},k)\,
\Gamma_{t\,s\,m'}^\sigma(k,\widetilde{k+q}) \nonumber\\
&&[\frac{n_t(k)[1-n_s(\widetilde{k+q})]}
{\omega+E_s(\widetilde{k+q})-E_t(k)-i\eta}
-\frac{n_s(\widetilde{k+q})[1-n_t(k)]}
{\omega+E_s(\widetilde{{k+q}})-E_t(k)+i\eta}]  \label{chilo}
\end{eqnarray}

and the vertex functions are given by

\begin{eqnarray}
&&\Gamma_{stm}^{\sigma}(\widetilde{k+q},k)=\sum_n A^{*}_{{\cal N},s}
(\widetilde{k+q})A_{n,t}(k){\sigma}^{n+{\cal N}}         \nonumber \\
&&\Gamma_{tsm}^{\sigma}(k,\widetilde{k+q})=(\Gamma_{stm}^{\sigma}(\widetilde{k+q},k))^*
\end{eqnarray}

Since we have to project $\chi_{m\,m'}^{\sigma\,\sigma'}(q,\omega)$
onto the charge and longitudinal spin sector we introduce the new matrices

\begin{equation}
\kappa_{mm'} = \frac{1}{2}\left( \!
\begin{array}{ c c}
\chi_{mm'}^{\rho \rho} & \chi_{mm'}^{\rho z}  \\
\chi_{mm'}^{z \rho}    & \chi_{mm'}^{z z}      \nonumber \\
\end{array}
\right)\,=\, A^{-1} \frac{1}{2}\left( \!
\begin{array}{ c c}
\chi_{mm'}^{\uparrow \uparrow}   & \chi_{mm'}^{\uparrow \downarrow}   \\
\chi_{mm'}^{\downarrow \uparrow} & \chi_{mm'}^{\downarrow \downarrow} \nonumber \\
\end{array}
\right)\,A.
\end{equation}

where 

\begin{equation}
A=\frac{1}{\sqrt{2}}\left( \!
\begin{array}{ c c}
1 &  \,\,\,\, 1   \\
1 &      -1 \nonumber \\
\end{array}
\right)
\end{equation}

Within these definitions we can rewrite eq. (\ref{sz}) as a
$2M \times 2M$ matrix equation for the charge and spin density
correlations respectively

\begin{equation}
\Xi\,=\,[1-U \Xi^o T_z]^{-1} \Xi^o
\end{equation}

and the matrices $T_z$ and $\Xi$ read as

\begin{equation}
\Xi \,=\, \left( \!
\begin{array}{ c c c c c }
\kappa_{11} & \kappa_{12} & \cdot & \cdot & \cdot  \\
\kappa_{21} & \kappa_{22} &       &       &   \\
\cdot       &             & \cdot &       &   \\
\cdot       &             &       & \cdot &   \\
\cdot       &             &       &       & \cdot  \nonumber \\
\end{array}
\right) \,\,;\,\, T_z \,=\, \left( \!
\begin{array}{ c c c c c }
\tau_z & 0      & 0      & \cdot & \cdot  \\
0      & \tau_z & 0      &       &        \\
0      & 0      & \tau_z &       &        \\
\cdot  &        &        & \cdot &        \\
\cdot  &        &        &       & \cdot  \nonumber \\
\end{array}
\right).
\end{equation}

Obviously both charge and longitudinal spin excitations are strongly
coupled except in the half-filled system where the ground state
corresponds to a commensurate antiferromagnet. 

\subsection{Effective interactions for general commensurability}

Having determined the collective excitations in the striped phase
we are now in the position to study the interactions of two holes
within the same subband with opposite spin and momenta in the spin- and 
charge channel respectively. Note that the susceptibilities defined in
eq. (\ref{SUS}) acquire non-diagonal matrix elements in the stripe state,
which implies the possibility of creating Cooper pairs with a finite momentum 
(which is a multiple of ${\bf Q^t}$). However, for simplicity we will 
consider only scattering of zero momentum Cooper pairs and therefore neglect nondiagonal contributions of the susceptibilities ($m \ne m'$) to the 
interaction. 
Additionally we will cast these interactions in the form 
of effective hamiltonians thus restricting on the static limit with an 
appropriate frequency cutoff. 

We find in the orientational spin-fluctuation channel 

\begin{eqnarray}\label{hh1}
H^{+-}&=&-\frac{U^2}{N}\sum_{k\,q \atop  m\,st}
\Gamma_{s\,t\,m}^{-}(\widetilde{k+q},k)\,
\Gamma_{s\,t\,m}^{+}(\widetilde{-k-q},-k)
\chi^{+-}_{mm}(q)  \nonumber\\
&\times& 
f^+_s(\widetilde{k+q},-\sigma)f^+_s(\widetilde{-k-q},\sigma)f_t(k,\sigma)f_t(-k,-\sigma)
\end{eqnarray}

in the charge-fluctuation channel

\begin{eqnarray}
H^{\rho \rho}&=&-\frac{U^2}{N}\sum_{k\,q \atop  m\,st}
\Gamma_{s\,t\,m}^\sigma(\widetilde{k+q},k)\,
\Gamma_{s\,t\,m}^{-\sigma}(\widetilde{-k-q},-k)
\chi^{\rho \rho}_{mm}(q)  \nonumber\\
&\times& 
f^+_s(\widetilde{k+q},-\sigma)f^+_s(\widetilde{-k-q},\sigma)f_t(k,\sigma)f_t(-k,-\sigma)
\end{eqnarray}

and in the amplitude spin-fluctuation channel

\begin{eqnarray}\label{hh2}
H^{zz}&=&-\frac{U^2}{N}\sum_{k\,q \atop  mm\,st}
\Gamma_{s\,t\,m}^\sigma(\widetilde{k+q},k)\,
\Gamma_{s\,t\,m}^{-\sigma}(\widetilde{-k-q},-k)
\chi^{zz}_{mm}(q)  \nonumber\\
&\times& 
f^+_s(\widetilde{k+q},-\sigma)f^+_s(\widetilde{-k-q},\sigma)f_t(k,\sigma)
f_t(-k,-\sigma).
\end{eqnarray}
 
In order to illustrate the behavior of the vertex functions in the
various channels let us consider the case of well developed stripe order
(i.e. large charge- and spin order parameters). 
In this limit we can as a first approximation neglect the kinetic energy 
in eq. (\ref{h1}) and the transformation eq. (\ref{Trans}) takes the form

\begin{equation}
A_{nt}(k) \approx \frac{1}{\sqrt{M}} exp [i\frac{2\pi}{M} nt].
\end{equation}

As a result the vertex contribution in the transverse spin channel
is given by 
\begin{equation}
\Gamma_{stm}^{-}(k,q)\Gamma_{stm}^{+}(k,q)=
(-1)^{(m-1)}\delta_{s,t+\frac{M}{2}}
\end{equation}
and thus the coupling to spin-flip scattering vanishes 
for intraband transitions. 

On the other hand the intraband coupling in the
charge and longitudinal spin channel survives \cite{foot2} and 
can be expressed in the form

\begin{equation}
\Gamma_{stm}^{\uparrow}(k,q)\Gamma_{stm}^{\downarrow}(k,q)=
(-1)^{(m-1)}\delta_{s,t}
\end{equation}

Now remember that the index {\it m} in $\chi_{mm}({\bf q})$ labels the 
multiple of ${\bf Q^{t}}$ which has to be added to the reduced transfered
momentum $\tilde{{\bf q}}$, i.e. the transfered momentum in the full BZ
is ${\bf q}=\tilde{\bf q}+(m-1){\bf Q^t}$. Thus an odd (even) index $m$
corresponds to enhanced charge (spin) susceptibilities which therefore 
in general mediate attractive (repulsive) interactions respectively.
In addition upon assuming that the charge (spin) susceptibility
has some finite contribution for odd (even) $m$ only, one recovers the 
result of Ref. \cite{Seibold} that the effective interaction in these
channels becomes proportional to the commensurability of the charge
modulation, i.e.
$\sum_m \Gamma_{stm}^{\uparrow}(k,q)\Gamma_{stm}^{\downarrow}(k,q) =\pm M/2$.

\section{Results}

We now apply the formalism developed above to the calculation of the 
effective interaction within the HF stripe state of the two-dimensional
Hubbard model. We restrict ourselves to the domain wall considered in
Sec. IIa, i.e. commensurability $M=16$ for which the total scattering vector
is $Q^t=(\pi,\pi+\frac{\pi}{8})$.
Fig. 5 displays the different RPA susceptibilities for $\omega=0$
and $\omega=25meV$ respectively. Note that the q-scans for the magnetic
susceptibilities are along the $(q_x,q_y=\pi)$ axis whereas the charge  
susceptibility is shown along $(q_x,q_y=0)$.
The structure in $\chi^{+-}$ is dominated by the Goldstone pole
at $Q^t$ due to the breaking of spin-rotational invariance in the
stripe state. As can be seen from Fig.6 the mode splits 
with increasing frequency  and the two branches
result from the defolding of the reduced BZ (Ref. \cite{KANESHITA}). 
Moreover beyond the gap energy (roughly corresponding to $\Delta_1$) the 
magnetic excitations rapidly lose intensity.

The longitudinal spin and charge channel do not display a physical pole at 
$\omega=0$. However, a mode like feature is present at $Q^t$ for the
amplitude fluctuations and at $2Q^t$ for the charge susceptibility which
reflects the enhanced susceptibility of the stripe
state to perturbations with the respective wave numbers.
Since both charge and spin are coupled within the RPA formalism the respective
excitations follow the same 
dispersion in ($q,\omega$)-space (Fig. 6) but
since the charge 'mode' appears as a second harmonic to the
longitudinal spin excitation its intensity is significantly lower.
It is interesting to observe that at the energy ($\approx \Delta_1$)
where the longitudinal spin fluctuations become commensurate the
charge fluctuations show the periodicity of the static spin modulation
(i.e. ${\bf q_s}$) which again is a consequence of the defolding of the reduced
BZ. Note that this energy does not correspond to the maximum
in the intensity, which appears for slightly lower frequencies. 

\begin{figure}
\begin{center}
\hspace{6cm}{{\psfig{figure=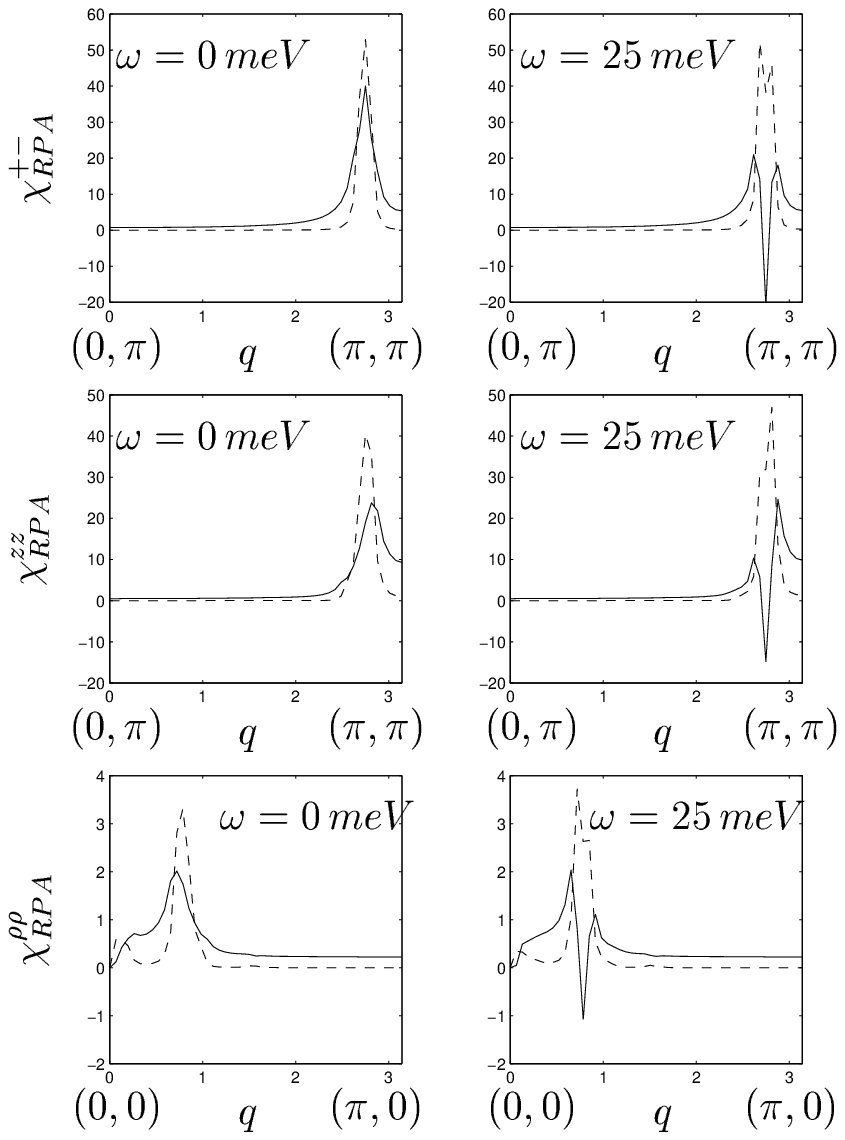,width=7cm}}}
\end{center}
\vspace*{0.2cm}
{\small FIG.5 The momentum dependence of the RPA susceptibilities
$\chi_{RPA}^{+-}$,$\chi_{RPA}^{zz}$,$\chi_{RPA}^{\rho \rho}$ 
in the stripe state for two frequences $\omega=0$ (left column)
and $\omega=25 meV$ (right column);
Solid: real part, Dashed: imaginary part; 
first and second lines represent the scans along path
$(0,\pi)-(\pi,\pi)$, third line  represents  the scan along path
$(0,0)-(\pi,0)$.}
\end{figure}

Finally, in order to determine the efficieny of the pairing in the
various channels we have to calculate the vertex functions which determine
the coupling of the charge carriers to the corresponding correlation
functions.
Generally speaking the vertex functions determine the overall sign of 
the effective interaction in the Brillouin zone. Let us briefly
illustrate this effect in the limit of a commensurate spin-density wave
($M=2$) where the vertex functions can be calculated analytically.
It is then shown below that the qualitative behavior of the coupling 
still holds for higher commensurability.
Note that the pairing interaction mediated by amplitude fluctuations of 
AF order constitutes the basis of the so called 
``spin bag mechanism to high-T$_c$ superconductivity''
as proposed by Schrieffer et al. in Ref. \cite{SCHRIEFF}.
In this case the antiferromagnetic scattering $Q=(\pi,\pi)$ doubles the unit
cell and we denote conduction and valence band by the superscripts 
$c$ and $v$ respectively. The corresponding energies are given
by $E^{c/v}_k=\pm E_k =\pm \sqrt{\Delta^2+\epsilon_k^2}$ and $\Delta$
is the SDW order parameter. 

Restricting ourselves on the 'hole doped' case the relevant coupling 
functions can be written within our notation as

\begin{eqnarray}
\Gamma_{v\,v\,m}^{\sigma}\Gamma_{v\,v\,m}^{-\sigma}&=&
\frac{(-1)^{m-1}}{2}\lbrack 1-\frac{(-1)^{m}\epsilon_k\epsilon_{k+q}-\Delta^2}{E_k E_{k+q}}
\rbrack \\
\Gamma_{v\,v\,m}^{+}\Gamma_{v\,v\,m}^{-}&=&
\frac{(-1)^{m-1}}{2}\lbrack 1-\frac{(-1)^{m}\epsilon_k\epsilon_{k+q}+\Delta^2}{E_k E_{k+q}}
\rbrack \label{trv1}
\end{eqnarray}

Obviously there is a difference in sign between small q- ($m=1$) and 
large q- ($m=2$) scattering. Since in an antiferromagnet the magnetic 
susceptibilities are peaked at $(\pi,\pi)$ the attractive interaction
for small momentum transfer (as anticipated in the spin-bag mechanism)
is accompanied by a large repulsion at ${\bf q} \sim {\bf Q}$ (see e.g.
Ref. \cite{CHUBUKOV},\cite{Kampf1} for a more detailed analysis of this point).

Let us now consider the case of commensurability M=16. As already mentioned
the considered hole doping $\delta=0.13$ is slightly above the 'magic' 
concentration $1/8$ so that the Fermi level is located within 
the 7th subband in the reduced BZ.

In Fig. 7 we have plotted the corresponding vertex functions for intraband
scattering which for convenience are defolded along $(q_x,0)$ and $(q_x,\pi)$.
It turns out that also in this case the coupling around odd multiples
of $Q^t$ is negativ whereas it is positive around the even harmonics.
Thus also in the stripe phase the large q scattering which is dominated
by the magnetic fluctuations is repulsive whereas attraction predominantly
arises from the coupling to charge fluctuations. Naturally for stripes 
the latter are much more pronounced in comparison to the spin-bag model.
With regard to the effective hamiltonian eqs. (\ref{hh1}-\ref{hh2}) the
cutoff of the pairing potential is an important parameter in determining
the interaction strength. From the right panel of Fig. 5 we 
conclude that for the relevant wave-vectors near the 'modes' the 
cutoff energy approximately is of the order 
of the charge order parameter $\chi_2$.

Besides the enhancement of charge fluctuation mediated pairing in the
stripe phase the contribution of spin-flip scattering to the interaction
is also significantly different from the spin-bag approach.
First note that the coupling to the 
transverse susceptibility for large momenta 
(upper right panel of Fig. (7)) is by two orders of magnitude 
smaller than the corresponding coupling to charge and longitudinal spin
fluctuations. In addition it vanishes at the momentum of the
Goldstone pole i.e. $q=Q^t$.
For the commensurate $M=2$ case this can be seen directly from 
eq. (\ref{trv1}) in the limit ${\bf q} \to (\pi,\pi)$ 
(note that in our reduced notation this is equivalent
to ${\bf q}=0$ in eq. (\ref{trv1}). In Ref. \cite{SCHRIEFF} this fact
led to some confusion concerning the relevance of the transverse fluctuations
within the spin-bag approach. However, it has been shown in Ref. 
\cite{FRENKEL} that the vanishing of the vertex is cancelled by the gapless
Goldstone pole of $\chi^{+-}(Q)$ and as a result the interaction in the 
spin-flip channel around $(\pi,\pi)$  becomes comparable to the pairing
from the amplitude fluctuations in the spin-bag scenario.

\begin{figure}
\begin{center}
\hspace{6cm}{{\psfig{figure=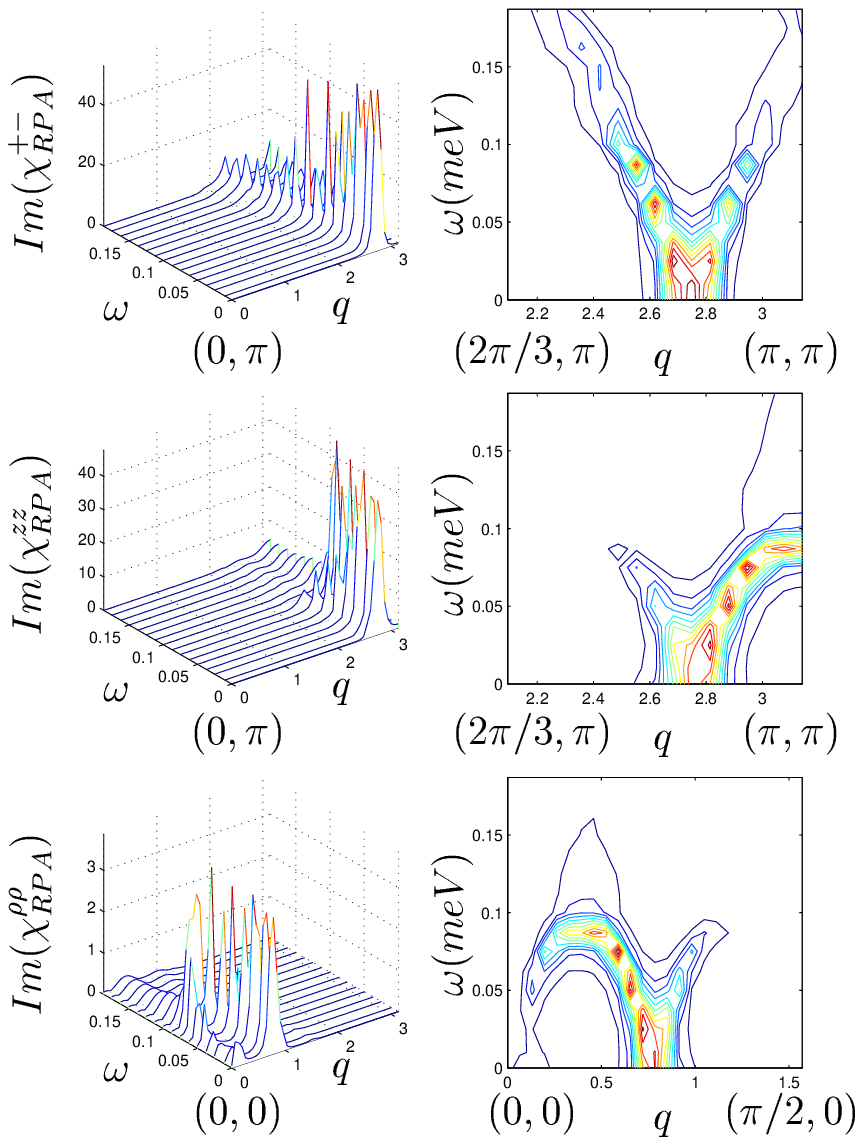,width=7cm}}}
\end{center}
\vspace*{0.2cm}
{\small FIG.6 The momentum and frequency dependences
of the imaginary parts of the RPA susceptibilities
$\chi_{RPA}^{+-}$,$\chi_{RPA}^{zz}$,$\chi_{RPA}^{\rho \rho}$ 
in the stripe state; right column shows the intensity
contour along $(0,\pi)-(\pi,\pi)$ for 
$Im(\chi_{RPA}^{+-})$,$Im(\chi_{RPA}^{zz})$ and
along $(0,0)-(\pi,0)$ for $Im(\chi_{RPA}^{\rho \rho})$..}
\end{figure}

In order to study the coupling to transverse fluctuations for commensurability
$M=16$ we have performed a numerical expansion of the vertex and the transverse
susceptibility around $q=Q^t$. 
Contrary to the low commensurability case
one finds that the contribution from the transverse channel to the
pairing is only $\approx 5 \%$ of that of the spin amplitude fluctuations.
We consider this as a consequence of the spatial separation of charge
carriers and spin fluctuations in the stripe model.

\section{Conclusion}

To summarize, we have calculated the effective interactions arising 
from the RPA
fluctuations around the HF stripe phase of the two-dimensional Hubbard model.
Since our approach is a generalization of Ref. \cite{SCHRIEFF} to higher 
commensurabilities the general features resemble closely those of the 
spin-bag mechanism.
However, significant differences arise due to the inhomogeneous charge 
distribution in the stripe state where the holes are confined to the domain 
walls separating antiphase AF regions. As a result the charge correlations 
are enhanced for wave-vectors
corresponding to the stripe periodicity resulting in a much more pronounced 
small q attractive pairing as compared to the spin-bag approach where the
system is translationally invariant with respect to the charge degrees of 
freedom.
Due to the fact that HF theory generates the charge order as a second 
harmonic of the longitudinal spin modulation the pairing is still dominated 
by the large q repulsion
due to magnetic fluctuations. However, in models where the stripe instability 
is driven by the charge an analogous approach should yield a comparable 
interaction strength in spin-and charge channels respectively.
Thus depending on the repulsive contribution due to magnetic scattering 
either a d-wave or anisotropic s-wave SC order parameter minimizes the 
free energy of the system. Obviously
non of these order parameters will be really realized in a purely 
one-dimensional stripe phase as considered in the present paper, however,
our investigations can be considered as complementary to Ref. 
\cite{MARTIN} where the coexistence between incommensurate antiferromagnetism
and anisotropic  superconductivity has been investigated.
In this context the effective interaction derived above may provide a 
microscopic
origin for the superconducting correlations considered in Ref. \cite{MARTIN}.  
Anyhow, in order to have true d-wave order one has 
to anticipate a scenario where the stripes fluctuate between the x- and y-
directions in order to restore the two-dimensional symmetry. 
In principle one could symmetrize the pairing interaction derived above
and use it in a BCS-type approach in order to evaluate the
superconducting gap. Alternatively, one could
consider the interactions arising from the fluctuations of an eggbox-type 
domain wall structure as in Ref. \cite{GOETZ} in order to mimic orientational 
stripe fluctuations. Furthermore it has been proposed \cite{NETO} that 
domains of stripes running along the crystalographic directions can
couple via the twin boundaries which also leads to 'symmetric' 
d-wave superconductivity.

\begin{figure}
\begin{center}
\hspace{6cm}{{\psfig{figure=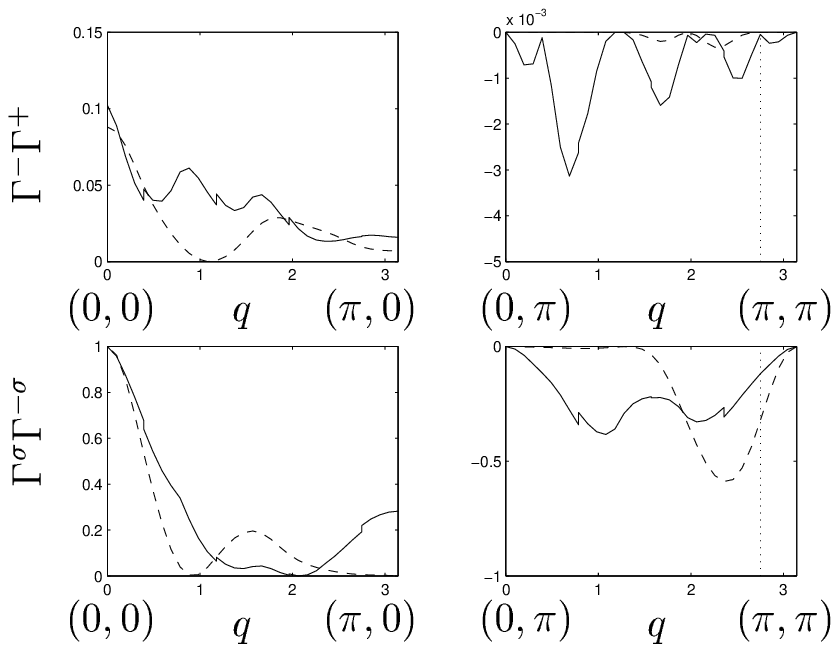,width=8cm}}}
\end{center}
\vspace*{0.2cm}
{\small FIG.7 The vertex functions 
$\Gamma_{77m}^-(k,q)\Gamma_{77m}^+(k,q)$ and
$\Gamma_{77m}^{\sigma}(k,q)\Gamma_{77m}^{-\sigma}(k,q)$ for two points 
in the reduced BZ:$k=(-\pi/1.6,-\pi/2.3)$, solid curve; $k=(0,0)$, 
dashed curve; The dotted line indicates the position of ${\bf Q^t}$.}
\end{figure}

However, within the obvious limitations of our approach
which we have already discussed in the introduction a quantitative
evaluation of the superconducting gap  would not
be very instructive and beyond the scope of the present paper.
Here our purpose is to simply
outline the general features of stripe fluctuation induced pairing
and to eventually supplement considerations within more sophisticated 
approaches.

\end{multicols}


\begin{references}
\bibitem{HAMMEL} P. C. Hammel, A. P. Reyes, Z. Fisk, M. Takigawa, J. D.
                Thompson, K. H. Heffner, and S. W. Chong, Phys. Rev.
                B {\bf 42}, 6781 (1990).
\bibitem{MESOT} J. Mesot, P. Allenspach, U. Staub, A. Furrer, and H. Mutka,
                Phys. Rev. Lett. {\bf 70}, 865 (1993).
\bibitem{TEITEL} G. B. Teitel'baum, B. B\"uchner, H.de Gronckel,
                  Phys. Rev. Lett. {\bf 84}, 2949 (2000).
\bibitem{BRINKMAN} A. Suter, M. Mali, J. Roos, D. Brinkmann,
                   J. Karpinski, and E. Kaldis, Phys. Rev. B {\bf 56}, 5542 (1997);
                   I. Eremin, M. Eremin, S. Varlamov, D. Brinkmann, M. Mali,
                   and J. Roos, Phys. Rev. B {\bf 56}, 11305 (1997). 
\bibitem{MEHRING} S. Kr\"amer, M. Mehring, Phys. Rev. Lett. {\bf 83},
                  396 (1999). 
\bibitem{HUNT} A. W. Hunt, P. M. Singer, K. R. Thurber, and T. Imai,
               Phys. Rev. Lett. {\bf 82}, 4300 (1999).
\bibitem{Zaanen1} J.Zaanen and O.Gunnarsson, Phys. Rev. B{\bf 40}, 7391 (1989).
\bibitem{Rice} D.Poilblanc and T.M.Rice, Phys. Rev. B{\bf 39}, 9739 (1989).
\bibitem{TRAN} J. M. Tranquada, B. J. Sternlieb, J. D. Axe,   
                Y. Nakamura, and S. Uchida, Nature {\bf 375}, 561 (1995).
                Y. Nakamura, and S. Uchida, Nature {\bf 375}, 561 (1995). 
\bibitem{NORMAND} B. Normand, A. P. Kampf, Phys. Rev. B {\bf 64}, 024521 
                  (2001).
\bibitem{YAM} K. Yamada, C. H. Lee, K. Kurahashi, J. Wada, S. Wakimoto,
                 S. Ueki, H. Kimura, Y. Endoh, S. Hosoya, G. Shirane,
                 R. J. Birgeneau, M. Greven, M. A. Kastner, and Y. J.Kim,
                 Phys. Rev. B {\bf 57}, 6165 (1998).
\bibitem{ARAI} M. Arai, T. Nishijima, Y. Endoh, T. Egami, S. Tajima, 
               K. Tomimoto, Y. Shiohara, M. Takahashi, A. Garrett, 
               S. Bennington, Phys. Rev. Lett. {\bf 83}, 608 (1999).
\bibitem{MOOK}H. A. Mook and F. Do\^gan, Nature {\bf 401}, 145 (1999).
\bibitem{BILLINGE} E.S. Bozin, G. H. Kwei, H. Takagi, and S. J. L. Billinge,
                    Phys. Rev. Lett. {\bf 84}, 5856 (2000).
\bibitem{ARAI2} M. Arai,  Proceedings 3rd International Conference on 
                'Stripes and high $T_c$ superconductivity', Roma (2000),
                will be published in Int. J. Mod. Phys. (2000).
\bibitem{CAST} C. Castellani, C. Di Castro, and M. Grilli, Phys. Rev. Lett.
               {\bf 75}, 4650 (1995); C. Castellani, C. Di Castro, 
               and M. Grilli, Z. Phys. B {\bf 103}, 137 (1997).
\bibitem{TALLON}J. L. Tallon, J. W. Loram, G. V. M. Williams, J. R. Cooper, 
                I. R. Fisher, J. D. Johnson, M. P. Staines, C. Bernhard, 
                Phys. Stat. Sol. B {\bf 215}, 531 (1999);
                J. L. Tallon, G. V. M. Williams, and J. W. Loram, 
                Physica C {\bf 338}, 9 (2000). 
\bibitem{BOEB} G. S. Boebinger, Yoichi Ando, A. Passner,T. Kimura, 
               M. Okuya, J. Shimoyama, K. Kishio, K. Tamasaku, N. Ichikawa,
               S. Uchida, Phys. Rev. Lett. {\bf 77}, 5417 (1996).
\bibitem{PERALI} A. Perali, C. Castellani, C. Di Castro, and M. Grilli,
        Phys. Rev. B {\bf 54}, 16216 (1996).
\bibitem{BALSEIRO} C. A. Balseiro and L. M. Falicov, Phys. Rev. B {\bf 20},
                   4457 (1979).
\bibitem{LAR} M.V.Eremin, I.A.Larionov, and S.V.Varlamov. Physica B 
                   {\bf 259-261}, 456 (1999).
\bibitem{SCHRIEFF} J.R.Schrieffer, X.G.Wen, and S.C.Zhang , 
                   Phys. Rev. B{\bf 39}, 11663 (1989).
\bibitem{FRENKEL} D.M.Frenkel and W.Hanke, Phys. Rev. B{\bf 42}, 6711 (1990).
\bibitem{ZAANEN} J. Zaanen and M. Ole\'{s}, Ann. Physik {\bf 5}, 224 (1996).
\bibitem{goetz1} G.Seibold, C.Castellani, C.Di Castro, M.Grilli, 
                   Phys.Rev. B{\bf 58}, 13506 (1998).
\bibitem{fleck} A. I. Lichtenstein, M. Fleck, A. M. Ole\'{s}, and L. Hedin,
                in {\it Stripes and Related Phenomena}, edited by
                A. Bianconi and N. L. Saini, Kluwer Academic (2000).
\bibitem{SCHULTZ} H.J.Schulz, Phys. Rev. L{\bf 64}, 1445 (1990);H.J.Schulz,
                  J.Phys.France {\bf 50}, 2833 (1989).
\bibitem{KANESHITA} E.Kaneshita, M.Ichioka and K.Machida, 
                    unpublished, cond-mat/0005466.
\bibitem{ZACHAR} O.Zachar, unpublished, cond-mat/0001217.
\bibitem{ARRIGONI} for a recent analysis of this scenario 
                   including fluctuation effects see e.g. 
                   E. Arrigoni and G.C.Strinati,Eur. Phys. J. B{\bf 19},
                   (2001) and references therein.
\bibitem{OVERHAUSER} A.W.Overhauser, Phys. Rev. {\bf 128}, 1437 (1962).
\bibitem{CHUBUKOV} A.V.Chubukov and D.M.Frenkel , Phys. Rev. B{\bf 46}, 11884 (1992).
\bibitem{INUI}  M.Inui and P.B.Littlewood, Phys. Rev. B {\bf 44}, 4415 (1991).
\bibitem{Zachar0} O.Zachar, S.A.Kivelson and V.J.Emery, Phys. Rev. B {\bf 57}, 
                  1422 (1998).
\bibitem{Seibold} G.Seibold and S.Varlamov, Phys. Rev. B{\bf 60}, 13056 (1999).
\bibitem{Kampf1} A.P.Kampf, Physics Reports {\bf 249}, 219 (1994).
\bibitem{MATSUDA}  M. Matsuda, M.Fujita, K.Yamada, R.J.Birgeneau, M.A.Kastner,
                   H.Hiraka, Y.Endoh, S.Wakimoto, G.Shirane, 
                   Phys. Rev. {\bf 62}, 9148 (2000). 
\bibitem{TRANQUADA} J. Tranquada, Physica B {\bf 241-243}, 745 (1997).
\bibitem{BLAIZOT} J.-P. Blaizot and G. Ripka, 
               {\it Quantum Theory of Finite Systems},
               (The MIT Press, 1986).
\bibitem{foot1} Since the maximum is the only extremum
                of the $(\langle H \rangle, \chi_2)$ curve the system cannot 
                acquire a self-consistent state with lower energy.
                Note that in this case the free energy is {\bf not} a 
                Landau functional which would have minima at finite
                order parameter values.  
\bibitem{foot2} Note that within our analysis the chemical potential 
                is located within
                a subband so that intraband excitations are still possible.
\bibitem{EMERY} V.J.Emery and S.A.Kivelson, Physica C {\bf 209}, 597 (1993).
\bibitem{MARTIN} I. Martin, G. Ortiz, A. V. Balatsky, and A. R. Bishop,
                 cond-mat/0009067.
\bibitem{GOETZ} G.Seibold, F.Becca, F.Bucci, C.Castellani, C.Di Castro, and 
                M.Grilli, Eur.Phys.J. B {\bf 513}, 87 (2000).
\bibitem{NETO} A. H. Castro Neto, cond-mat/0102281.

\end{references}
\end{document}